\documentclass[12pt]{article}
\usepackage{amsfonts}
\usepackage{amsmath}
\usepackage{amssymb}
\usepackage{myart}
\usepackage[dvips]{graphicx}

\normalbaselines
\font\tenbm=cmmib10
\font\sevenbm=cmmib7

\newfam\bmfam
\textfont\bmfam=\tenbm \scriptfont\bmfam=\sevenbm
\input tcilatex.tex
\begin{document}

\author{Yuri A. Rylov}
\title{Discriminating properties of compactification in discrete uniform
isotropic space-time}
\date{Institute for Problems in Mechanics, Russian Academy of Sciences \\
101-1 ,Vernadskii Ave., Moscow, 117526, Russia \\
email: rylov@ipmnet.ru\\
Web site: {$http://rsfq1.physics.sunysb.edu/\symbol{126}rylov/yrylov.htm$}\\
or mirror Web site: {$http://gasdyn-ipm.ipmnet.ru/\symbol{126}%
rylov/yrylov.htm$}}
\maketitle

\begin{abstract}
Compactification of the 5-dimensional Kaluza-Klein space-time
geometry is considered. The space-time geometry is supposed to be
discrete, uniform and isotropic. It is shown, that consideration
of the space-time geometry as a physical geometry, i.e. as a
geometry described completely by the single-valued world function,
leads to a discrimination of some values of the particle charge.
At the conventional approach, when the world function becomes to
be many-valued after compactification, the value of the elementary
particle electric charge remains to be unrestricted, and this fact
does not agree with experimental data. It is important, that the
discrete geometry is given on the continual set of points. This
circumstance makes admissible a compatibility of discreteness with
the uniformity and isotropy of the geometry.
\end{abstract}

\section{Introduction}

The role of space-time geometry in description of physical phenomena of
microcosm has been increased due to appearance of a more general conception
of geometry. In the twentieth century the Riemannian geometry was considered
to be the most general geometry, suitable for description of the space-time.
However, the Riemannian geometry cannot describe such properties of
space-time as discreteness, restricted divisibility of geometrical objects
and discrete characteristics (mass, charge, angular momentum) of elementary
particles. Discrete characteristics of elementary particles are considered
usually to be dynamic properties of elementary particles.

In reality, at a use of a true conception of the space-time geometry the
elementary particles in themselves, as well as their properties and their
dynamics can be described in terms of the proper space-time geometry and
only in terms of the space-time geometry. The conventional conception of
geometry, which supposes, that any geometry is axiomatizable, and any
geometry can be deduced from a system of axioms, is wrong. In any
axiomatizable geometry the equivalence relation is supposed to be
transitive. Only at the transitive equivalence relation the set of all
geometric propositions (i.e. geometry) can be deduced from an axiomatics (a
finite set of basic geometric propositions).

A new method of the physical geometry construction has been invented in the
end of the twentieth century \cite{R2001}. The physical geometry is such a
geometry, which is described completely by the world function $\sigma $. The
world function $\sigma \left( P,Q\right) $ is a single-valued real function
of any two points $P,Q\in \Omega $, where $\Omega $ is the set of all points
(or events), where the geometry is given.%
\begin{equation}
\sigma :\qquad \Omega \times \Omega \rightarrow \mathbb{R},\qquad \sigma
\left( P,P\right) =0,\qquad \forall P\in \Omega  \label{a1.1}
\end{equation}%
The world function $\sigma \left( P,Q\right) =\frac{1}{2}\rho ^{2}\left(
P,Q\right) $, where $\rho \left( P,Q\right) $ is the distance between the
points $P$ and $Q$.

On one hand, the proper Euclidean geometry $\mathcal{G}_{\mathrm{E}}$ is the
axiomatizable geometry, which can be deduced from the Euclidean axiomatics
\cite{H30}. On the other hand, the proper Euclidean geometry $\mathcal{G}_{%
\mathrm{E}}$ is a physical geometry. It means, that all definitions $%
\mathcal{D}_{\mathrm{E}}$ of $\mathcal{G}_{\mathrm{E}}$ can be expressed in
terms of the Euclidean world function in the form $\mathcal{D}_{\mathrm{E}}=%
\mathcal{D}_{\mathrm{E}}\left[ \sigma _{\mathrm{E}}\right] $. There is a
theorem, where this statement has been proved \cite{R1990,R2001}. If now one
replaces the Euclidean world function $\sigma _{\mathrm{E}}$ with the world
function $\sigma $ of some other physical geometry $\mathcal{G}$ in all
definitions $\mathcal{D}_{\mathrm{E}}:\mathcal{D}_{\mathrm{E}}\left[ \sigma
_{\mathrm{E}}\right] \rightarrow \mathcal{D}_{\mathrm{E}}\left[ \sigma %
\right] $, one obtains all definitions $\mathcal{D}_{\mathrm{E}}\left[
\sigma \right] $ of the physical geometry $\mathcal{G}$. The procedure of
replacement is a deformation of the proper Euclidean geometry, when the
Euclidean distance $\rho _{\mathrm{E}}=\sqrt{2\sigma _{\mathrm{E}}}$ are
replaced by the distance $\rho =\sqrt{2\sigma }$ of the physical geometry $%
\mathcal{G}$. Thus, any physical geometry is obtained from the proper
Euclidean geometry by means of a deformation.

In general, the physical geometry is not axiomatizable, because the
axiomatizability of a geometry is possible only, if the equivalence relation
is transitive. Indeed, in the proper Euclidean geometry the vector $\mathbf{P%
}_{0}\mathbf{P}_{1}\ $is defined as an ordered set $\mathbf{P}_{0}\mathbf{P}%
_{1}=\left\{ P_{0},P_{1}\right\} $ of two points $P_{0},$ $P_{1}$. The
equivalence (equality) of two vectors$\ \mathbf{P}_{0}\mathbf{P}_{1}$ and $%
\mathbf{Q}_{0}\mathbf{Q}_{1}$ is defined by two relations. Two vectors $%
\mathbf{P}_{0}\mathbf{P}_{1}$ and $\mathbf{Q}_{0}\mathbf{Q}_{1}$ are
equivalent ($\mathbf{P}_{0}\mathbf{P}_{1}\mathrm{eqv}\mathbf{Q}_{0}\mathbf{Q}%
_{1}$), if%
\begin{equation}
\mathbf{P}_{0}\mathbf{P}_{1}\mathrm{eqv}\mathbf{Q}_{0}\mathbf{Q}_{1}:\qquad
\left( \mathbf{P}_{0}\mathbf{P}_{1}.\mathbf{Q}_{0}\mathbf{Q}_{1}\right)
=\left\vert \mathbf{P}_{0}\mathbf{P}_{1}\right\vert \cdot \left\vert \mathbf{%
Q}_{0}\mathbf{Q}_{1}\right\vert \wedge \left\vert \mathbf{P}_{0}\mathbf{P}%
_{1}\right\vert =\left\vert \mathbf{Q}_{0}\mathbf{Q}_{1}\right\vert
\label{a1.2}
\end{equation}%
where the scalar product $\left( \mathbf{P}_{0}\mathbf{P}_{1}.\mathbf{Q}_{0}%
\mathbf{Q}_{1}\right) $ of vectors $\mathbf{P}_{0}\mathbf{P}_{1}$ and $%
\mathbf{Q}_{0}\mathbf{Q}_{1}$ is defined by the relation%
\begin{equation}
\left( \mathbf{P}_{0}\mathbf{P}_{1}.\mathbf{Q}_{0}\mathbf{Q}_{1}\right)
=\sigma \left( P_{0},Q_{1}\right) +\sigma \left( P_{1},Q_{0}\right) -\sigma
\left( P_{0},Q_{0}\right) -\sigma \left( P_{1},Q_{1}\right) ,\quad \forall
P_{0},P_{1},Q_{0},Q_{1}\in \Omega  \label{a1.3}
\end{equation}%
\begin{equation}
\left\vert \mathbf{P}_{0}\mathbf{P}_{1}\right\vert ^{2}=2\sigma \left(
P_{0},P_{1}\right)  \label{a1.3a}
\end{equation}%
and $\sigma $ means the world function of the proper Euclidean geometry. The
first relation of (\ref{a1.2}) describes parallelism of vectors $\mathbf{P}%
_{0}\mathbf{P}_{1}$ and $\mathbf{Q}_{0}\mathbf{Q}_{1}$, whereas the second
one describes equality of their lengths. The definition of equivalence of
two vectors contains only points $P_{0},P_{1},Q_{0},Q_{1}$, determining the
vectors, and world functions between these points. The definition does not
refer to a coordinate system and to the dimension of the proper Euclidean
geometry $\mathcal{G}_{\mathrm{E}}$. It is a pure geometric definition,
which does not contain a reference to the means of description. In the
proper Euclidean geometry the definition (\ref{a1.2}) of equivalence
coincides with the conventional equivalence definition on the ground of the
linear vector space. The equivalence relation (\ref{a1.2}) is transitive in
the proper Euclidean geometry $\mathcal{G}_{\mathrm{E}}$, and this
transitivity is a special property of the proper Euclidean geometry.

In the arbitrary physical geometry $\mathcal{G}$ the definition of
equivalence has the same form (\ref{a1.2}) with the world function $\sigma $%
, describing the geometry $\mathcal{G}$. However, in the general case the
equivalence relation (\ref{a1.2}) is not transitive, in general, because in
the case of arbitrary world function $\sigma $ the equivalence of two
vectors is multivariant, in general. It means that at the point $P_{0}$ may
exist many vectors $\mathbf{P}_{0}\mathbf{P}_{1}$, $\mathbf{P}_{0}\mathbf{P}%
_{1}^{\prime }$, $\mathbf{P}_{0}\mathbf{P}_{1}^{\prime \prime },...,$ which
are equivalent to the vector $\mathbf{Q}_{0}\mathbf{Q}_{1}$ at the point $%
Q_{0}$, whereas the vectors $\mathbf{P}_{0}\mathbf{P}_{1}$, $\mathbf{P}_{0}%
\mathbf{P}_{1}^{\prime }$, $\mathbf{P}_{0}\mathbf{P}_{1}^{\prime \prime
},... $are not equivalent between themselves. In this case it is possible,
that
\begin{equation}
\mathbf{P}_{0}\mathbf{P}_{1}\mathrm{eqv}\mathbf{Q}_{0}\mathbf{Q}_{1}\wedge
\mathbf{P}_{0}\mathbf{P}_{1}^{\prime }\mathrm{eqv}\mathbf{Q}_{0}\mathbf{Q}%
_{1}\wedge \mathbf{P}_{0}\mathbf{P}_{1}\overline{\mathrm{eqv}}\mathbf{P}_{0}%
\mathbf{P}_{1}^{\prime }  \label{a1.4}
\end{equation}%
is true. Here the symbol $\overline{\mathrm{eqv}}$ means non-equivalency. If
relations (\ref{a1.4}) take place, the equivalence relation is intransitive,
because for transitive equivalence relation it follows from
\begin{equation}
\mathbf{P}_{0}\mathbf{P}_{1}\mathrm{eqv}\mathbf{Q}_{0}\mathbf{Q}_{1}\wedge
\mathbf{P}_{0}\mathbf{P}_{1}^{\prime }\mathrm{eqv}\mathbf{Q}_{0}\mathbf{Q}%
_{1}  \label{a1.5}
\end{equation}%
that%
\begin{equation}
\mathbf{P}_{0}\mathbf{P}_{1}\mathrm{eqv}\mathbf{P}_{0}\mathbf{P}_{1}^{\prime
}  \label{a1.6}
\end{equation}%
and the relation (\ref{a1.4}) is false. On the other hand, the number of
vectors $\mathbf{P}_{0}\mathbf{P}_{1}$, $\mathbf{P}_{0}\mathbf{P}%
_{1}^{\prime }$, $\mathbf{P}_{0}\mathbf{P}_{1}^{\prime \prime },...,$ which
are equivalent to the vector $\mathbf{Q}_{0}\mathbf{Q}_{1}$ at the point $%
Q_{0}$ depends on the number of solutions of two equations (\ref{a1.2}),
considered as equations for determination of the point $P_{1}$ at fixed
points $P_{0},Q_{0},Q_{1}$ (or at fixed point $P_{0}$ and fixed vector $%
\mathbf{Q}_{0}\mathbf{Q}_{1}$). The number of these solutions depends on the
form of the world function $\sigma $. We shall differ three different cases:

(1) Single-variance with respect to points $P_{0},Q_{0},Q_{1}$, when there
is one and only one solution for $P_{1}$ at given points $P_{0},Q_{0},Q_{1}$%
. In this case the equivalence relation is transitive, if the
single-variance takes place for any points $P_{0},Q_{0},Q_{1}$.

(2) Multivariance with respect to points $P_{0},Q_{0},Q_{1}$, when there is
more, than one solution at some given points $P_{0},Q_{0},Q_{1}$. In this
case the equivalence relation is intransitive.

(3) zero-variance with respect to points $P_{0},Q_{0},Q_{1}$, when there is
no solution at some given points $P_{0},Q_{0},Q_{1}$. In this case the
equivalence relation may be intransitive and may be transitive.

The second case is strongest in the sense, that appearance of multivariance
with respect to some three points $P_{0},Q_{0},Q_{1}$ generates
intransitivity of the equivalence relation, and, hence, non-axiomatizability
of a physical geometry, because the transitivity of the equivalence relation
is a necessary condition of axiomatizability. The second case and the third
one are compatible in the sense that the multivariance may take place with
respect to some points $P_{0},Q_{0},Q_{1}$, whereas the zero-variance may
take place with respect to other points $P_{0}^{\prime },Q_{0}^{\prime
},Q_{1}^{\prime }$.

Note that the geometry of Minkowski may be axiomatizable and non-physical,
and the geometry of Minkowski may be physical and nonaxiomatizable. In
general, in this case one has two different geometries, having the same
world function. We use for them different names. The geometry of Minkowski,
which is a physical geometry, will be referred to as the $\sigma $%
-Minkowskian geometry. The $\sigma $-Minkowskian geometry is not
axiomatizable, because it is multivariant with respect to any point $P_{0}$
and any spacelike vector $\mathbf{Q}_{0}\mathbf{Q}_{1}=\left\{
Q_{0},Q_{1}\right\} $. The conventional geometry of Minkowski, which is
constructed on the ground of the linear vector space with the scalar
product, given on it, is axiomatizable (it is deduced from some axiomatics),
but it is not a physical geometry. The geometry of Minkowski cannot be
constructed on the basis of the world function only. Construction of the
geometry of Minkowski contains a reference to the means of description in
the form of the coordinate system. Although the construction of the geometry
of Minkowski is invariant with respect to transformation of the coordinate
system, it is not invariant with respect to transformation of the coordinate
system dimension (see detailed discussion in \cite{R2008}). The geometry of
Minkowski should be qualified as a fortified physical geometry, i.e. a
physical geometry with some additional structure, given on the physical
geometry. Existence of the additional structure imposes some additional
constraints on the geometry.

The difference between the space-time geometry of Minkowski and the $\sigma $%
-Minkow\-skian space-time geometry appears only at consideration of
spacelike vectors, with respect to which the $\sigma $-Minkowskian geometry
is multivariant. However, the spacelike vectors do not figure in the
particle dynamics, and the difference between the $\sigma $-Minkowskian
space-time geometry and the space-time geometry of Minkowski remains to be
obscure. If one considers the geometry as a science on mutual disposition of
geometrical objects and their shapes, one should prefer the $\sigma $%
-Minkowskian geometry as a space-time geometry, because the distance between
any pair of points determines mutual disposition of geometrical objects and
their shapes completely. As to axiomatizability of a geometry, this property
is important only for deduction of the geometric propositions from the
axiomatics. From viewpoint of the geometry as a science on disposition of
geometrical object, the axiomatizability is a secondary property of the
geometry, and practically all physical geometries are not axiomatizable. The
proper Euclidean geometry is a very important exclusion, which admits one to
construct physical geometries by means of a deformation of the proper
Euclidean geometry.

Deduction of an axiomatizable geometry from axiomatics has two essential
defects. Firstly, one needs to formulate geometric propositions and to prove
corresponding theorems. The geometric propositions are to be formulated and
proved for any new geometry. These procedures are complicated from the
technical viewpoint. Besides, only geometries with the transitive
equivalence relation can be deduced from axiomatics. Secondly, one needs to
invent axioms, and to test their consistency. Inconsistency of a geometry
means, that using two different ways of deduction of some statement, one
obtains two incompatible statements. \textit{Inconsistency of a geometry is
a property of the method of the geometry construction, but not a property of
the geometry in itself.} In the physical geometry, which is constructed on
the ground of the deformation principle, the question of its inconsistency
is meaningless, because the problem of geometric propositions formulation is
absent at all. All definitions of a physical geometry are taken from the
proper Euclidean geometry in the ready-made form. More exactly, definitions
of geometrical objects are taken from the proper Euclidean geometry. If it
is necessary to determine properties of these objects, they are calculated
on the basis of the world function. As far as the world functions are
different, in general, in the considered geometry and in the Euclidean one,
the calculated properties may be different in the considered geometry and in
the Euclidean one.

Finally, the method of a physical geometry construction, based on linear
vector space, (for instance, construction of the Riemannian geometry) starts
from some $n$-dimensional manifold $\mathcal{M}_{n}$, where the metric
tensor $g_{ik}$ is given. The world function $\sigma $ is given by the
relation
\begin{equation}
\sigma \left( x,x^{\prime }\right) =\frac{1}{2}\left( \dint\limits_{\mathcal{%
L}_{xx^{\prime }}}\sqrt{g_{ik}\left( x\right) dx^{i}dx^{k}}\right) ^{2}
\label{a1.7}
\end{equation}%
where integration is produced along the geodesic $\mathcal{L}_{xx^{\prime }}$%
, connecting points $x$ and $x^{\prime }$. There may be several geodesics,
connecting points $x$ and $x^{\prime }$. In this case the world function $%
\sigma $ appears to be many-valued. In this case the world function is a
derivative quantity, and it may be many-valued. However, in a physical
geometry, the world function is a primary quantity, it determines the
physical geometry, and it cannot be many-valued.

To make the Riemannian geometry with many-valued world function (\ref{a1.7})
a physical geometry ($\sigma $-Riemannian geometry), one needs to turn the
many-valued world function into single-valued one, choosing only one branch
of the world function (\ref{a1.7}). Different choice of branches generates
different world functions and, hence, different $\sigma $-Riemannian
geometries. Thus, the $n$-dimensional manifold $\mathcal{M}_{n}$ with the
metric tensor, given on it generates several $\sigma $-Riemannian
geometries, if the expression (\ref{a1.7}) appears to be many-valued for
some pairs of points $x,x^{\prime }$.

Construction of $\sigma $-Riemannian geometries by means of a transformation
of many-valued world function (\ref{a1.7}) into a single-valued world
function is accompanied by appearance of zero-variance for some points. Of
course, this mechanism of construction of a physical geometry with the
zero-variance is not unique. However, this mechanism is interesting from the
physical viewpoint, because the $\sigma $-Riemannian geometry with the
zero-variance may be obtained as a result of the compactification of the
flat space-time geometry (for instance, compactification of 5-dimensional
space-time geometry of Kaluza-Klein \cite{K1926,V2005}). The zero-variance
generates some discrimination mechanism, responsible for discrete values of
the elementary particle parameters. In particular, compactification of the
fifth coordinate in the Kaluza-Klein geometry leads to restrictions on the
possible electric charge of the elementary particle.

This paper is devoted to consideration of the procedure of compactification
of the Kaluza-Klein geometry, which is accompanied by the construction of a
discrimination mechanism, imposing restrictions on the value of the electric
charge of the elementary particles. However, at first, we mention about
influence of the multivariance upon the dynamics of elementary particles.

\section{Influence of the multivariance on the particle dynamics.}

In the space-time geometry of Minkowski the dynamics of a pointlike particle
is described by a timelike world line $\mathcal{L}$ of the particle. In the
inertial coordinate system $x=\left\{ x^{0},x^{1},x^{2},x^{3}\right\} $ the
world function $\sigma _{\mathrm{M}}\left( x,x^{\prime }\right) $ between
two points with coordinates $x$ and $x^{\prime }$ has the form

\begin{equation}
\sigma _{\mathrm{M}}\left( x,x^{\prime }\right) =\frac{1}{2}g_{ik}\left(
x^{i}-x^{\prime i}\right) \left( x^{k}-x^{\prime k}\right)  \label{a3.1}
\end{equation}%
where the metric tensor has the form $g_{ik}=\mathrm{diag}\left\{
c^{2},-1,-1,-1\right\} $, and $c$ is the speed of the light. The world line $%
x=x\left( \tau \right) $ of a charged particle, moving in the given
electromagnetic field $F_{ik}$, is described by the dynamic equation%
\begin{equation}
m\frac{d}{d\tau }\frac{cg_{il}\frac{dx^{l}}{d\tau }}{\sqrt{g_{jn}\frac{dx^{j}%
}{d\tau }\frac{dx^{n}}{d\tau }}}=\frac{e}{c}F_{ik}\left( x\right) \frac{%
dx^{k}}{d\tau },\qquad i=0,1,2,3  \label{a3.2}
\end{equation}%
where $m$ is the particle mass, $e$ is the electric charge of the particle
and $\tau $ is a parameter along the world line. The constants $\ m$ and $e$
are non-geometrical characteristics of the pointlike particle.

In general, the mass $m$ and the charge $e$ can be geometrized, i.e. they
may be considered as pure geometric characteristics of the pointlike
particle. However, it is possible only in the framework of the physical
geometry, which is formulated in terms of the world function. The motion of
a pointlike particle is described by a world chain $\mathcal{C}$, consisting
of connected vectors $\mathbf{P}_{s}\mathbf{P}_{s+1}$, $s=...0,1,...$%
\begin{equation}
\mathcal{C=}\dbigcup\limits_{s}\mathbf{P}_{s}\mathbf{P}_{s+1}  \label{a3.2a}
\end{equation}%
where vector $\mathbf{P}_{s}\mathbf{P}_{s+1}=\left\{ P_{s},P_{s+1}\right\} $
is an ordered set of two points $P_{s},P_{s+1}$. The world chain $\mathcal{C}
$ is an ordered set of points $...P_{0},P_{1},...P_{s},...$ The distance $%
\left\vert \mathbf{P}_{s}\mathbf{P}_{s+1}\right\vert $ between the adjacent
points $P_{s},P_{s+1}$ is the same.%
\begin{equation}
\left\vert \mathbf{P}_{s}\mathbf{P}_{s+1}\right\vert =\mu ,\qquad s=...0,1,..
\label{a3.3}
\end{equation}%
The quantity $\mu =\left\vert \mathbf{P}_{s}\mathbf{P}_{s+1}\right\vert $ is
the length of the world chain link. It determines the geometric mass $\mu $.
The geometrical mass $\mu $ is connected with the usual mass $m$ by means of
the relation%
\begin{equation}
m=b\mu  \label{a3.4}
\end{equation}%
where $b$ is some universal constant. The geometrical mass $\mu $ is a
geometric characteristic of the particle, as well as the vector $\mathbf{P}%
_{s}\mathbf{P}_{s+1}$, which is the geometric momentum of the particle.

The motion (\ref{a3.2}) of a pointlike particle in the electromagnetic field
may be described as a free motion of the particle in the 5-dimensional
space-time of Kaluza-Klein. The fact, that the motion of a pointlike
particle in a physical space-time geometry is free, means that the adjacent
vectors in the world chain are equivalent
\begin{equation}
\mathbf{P}_{s}\mathbf{P}_{s+1}\mathrm{eqv}\mathbf{P}_{s+1}\mathbf{P}%
_{s+2},\qquad s=...0,1,..  \label{a3.5}
\end{equation}

Let the electromagnetic field be absent. Then dynamic equation (\ref{a3.2})
turns into the dynamic equation
\begin{equation}
m\frac{d}{d\tau }\frac{cg_{il}\frac{dx^{l}}{d\tau }}{\sqrt{g_{jn}\frac{dx^{j}%
}{d\tau }\frac{dx^{n}}{d\tau }}}=0  \label{a3.6}
\end{equation}%
Its solution%
\begin{equation}
x^{i}=x^{i}\left( \tau \right) =X^{i}+U^{i}\tau ,\qquad X^{i},U^{i}=\text{%
const},\qquad i=0,1,2,3  \label{a3.7}
\end{equation}%
does not depend on the mass $m$ and coincides with the solution of equations
(\ref{a3.5}), (\ref{a1.2}) in the space-time of Minkowski. However, if the
space-time of Minkowski is slightly deformed, the solution may appear to
depend on the mass.

Let us consider the space-time geometry $\mathcal{G}_{\mathrm{d}}$,
described by the world function%
\begin{equation}
\sigma _{\mathrm{d}}=\sigma _{\mathrm{M}}+d,\qquad d=\frac{1}{2}\lambda
_{0}^{2}\mathrm{sgn}\left( \sigma _{\mathrm{M}}\right) ,\qquad \lambda _{0}=%
\text{const}  \label{a3.8}
\end{equation}%
\begin{equation}
\mathrm{sgn}\left( x\right) =\left\{
\begin{array}{lll}
\frac{x}{\left\vert x\right\vert } & \text{if} & x\neq 0 \\
0 & \text{if} & x=0%
\end{array}%
\right.  \label{a3.8a}
\end{equation}%
where $\sigma _{\mathrm{M}}$ is the world function of the space-time
geometry of Minkowski, and $\lambda _{0}$ is some elementary length of the
geometry $\mathcal{G}_{\mathrm{d}}$.

The length $\left\vert \mathbf{P}_{0}\mathbf{P}_{1}\right\vert _{\mathrm{d}}$
of any vector $\mathbf{P}_{0}\mathbf{P}_{1}$ has the form%
\begin{equation}
\left\vert \mathbf{P}_{0}\mathbf{P}_{1}\right\vert _{\mathrm{d}}^{2}=2\sigma
_{\mathrm{d}}\left( P_{0},P_{1}\right) =2\sigma _{\mathrm{M}}\left(
P_{0},P_{1}\right) +\lambda _{0}^{2}\mathrm{sgn}\left( \sigma _{\mathrm{M}%
}\left( P_{0}.P_{1}\right) \right)  \label{a3.9}
\end{equation}

In other words, if the distance between points $P_{0},P_{1}$ is timelike in $%
\mathcal{G}_{\mathrm{d}}$ $\left( \sigma _{\mathrm{d}}\left(
P_{0},P_{1}\right) >0\right) $, it is also timelike in $\mathcal{G}_{\mathrm{%
M}}$ $\left( \sigma _{\mathrm{M}}\left( P_{0},P_{1}\right) >0\right) $. If
the distance between points $P_{0},P_{1}$ is spacelike in $\mathcal{G}_{%
\mathrm{d}}$ $\left( \sigma _{\mathrm{d}}\left( P_{0},P_{1}\right) <0\right)
$, it is also spacelike in $\mathcal{G}_{\mathrm{M}}$ $\left( \sigma _{%
\mathrm{M}}\left( P_{0},P_{1}\right) <0\right) $. It follows from (\ref{a3.9}%
) that any timelike (and spacelike) distance is larger, than $\lambda _{0}$.
It means that in the space-time geometry $\mathcal{G}_{\mathrm{d}}$ there
are no close points, and the geometry $\mathcal{G}_{\mathrm{d}}$ should be
qualified as a discrete space-time geometry. The geometry $\mathcal{G}_{%
\mathrm{d}}$ is given on the continuous manifold of Minkowski. It looks
rather unexpected, that the discrete geometry may be given on the same point
set, on which a continuous geometry can be given. This surprise is explained
by the fact, that at the conventional approach, based on the concept of the
linear space, the discrete geometry is given on a countable point set,
whereas the continuous geometry is given on a continual point set.

\label{7b}Conventionally a discrete geometry is described as follows. Let us
consider some geometry $\mathcal{G}_{\mathrm{c}}$ (Euclidean, Minkowskian,
or Riemannian) on some manifold $\mathbb{M}_{n}$ and introduce some
curvilinear coordinate system $\left( x^{0},x^{1},...x^{n}\right) $ in it.
Let us remove from the manifold $\mathbb{M}_{n+1}$ all points except of
points with all integer coordinates. As a result one obtains the point set $%
\mathbb{M}_{\mathrm{d}}$, whose points a labelled by integer coordinates $%
x^{s}$, $s=0,1,,..n$. The world function $\sigma (P,Q)$ between the points $%
P,Q\in \mathbb{M}_{\mathrm{d}}$ is the same as between the corresponding
points $P,Q\in \mathbb{M}_{n+1}$. As a result one obtains the same geometry $%
\mathcal{G}_{\mathrm{c}}$ on the subset $\mathbb{M}_{\mathrm{d}}$ of the set
$\mathbb{M}_{n+1}$. In the discrete geometry $\mathcal{G}_{\mathrm{c}}$
defined on $\mathbb{M}_{\mathrm{d}}$ there is an elementary length $\lambda $%
, defined by the relation%
\begin{equation}
\lambda =\min_{\forall P,Q\in \mathbb{M}_{\mathrm{d}}}\left\{ \left\vert
\sqrt{2\sigma \left( P,Q\right) }\right\vert \right\} \ \ \ \text{at }%
\left\vert \sqrt{2\sigma \left( P,Q\right) }\right\vert >0  \label{a3.9a}
\end{equation}%
In this conventional definition of the discrete geometry one uses such means
of description as the manifold $\mathbb{M}_{n+1}$ and a coordinate system on
it. The obtained geometry on $\mathbb{M}_{\mathrm{d}}$ depends essentially
on the choice of the coordinate system. Besides, it is impossible to obtain
a discrete geometry on a continuous set of points.

The definition (\ref{a3.8}) does not use any means of description. It uses
only world function, and discreteness of the geometry arises from the fact
that $\left\vert \sigma \left( P,Q\right) \right\vert \notin \left(
0,\lambda _{0}\right) $ for $\forall P,Q\in \Omega $, where $\Omega $ is the
point set where the geometry is given. The set $\Omega $ may be discrete or
continuous. This circumstance is unessential for construction of the
discrete geometry.\label{7e}

The character (discreteness, or continuity) of geometry depends only on the
form of the world function. Of course, the continual geometry may be given
only on a continual point set. However, as we have seen, the discrete
geometry may be given also on a continual point set .

As we have seen, the $\sigma $-Minkowskian geometry is multivariant with
respect to any point and any spacelike vector. The space-time geometry $%
\mathcal{G}_{\mathrm{d}}$ is multivariant with respect to timelike vectors
also, and this circumstance appears to be important for dynamics of a
pointlike particle, because the dynamics deals with timelike vectors. The
free motion of a pointlike particle appears to depend on the geometric
particle mass $\mu $ and on the elementary length $\lambda _{0}$, which is
responsible for multivariance of $\mathcal{G}_{\mathrm{d}}$ with respect to
timelike vectors.

Two adjacent links $\mathbf{P}_{0}\mathbf{P}_{1}$ and $\mathbf{P}_{1}\mathbf{%
P}_{2}$ are equivalent, and, hence, satisfy the relations of the type of (%
\ref{a1.2}). Let coordinates of the points be
\begin{equation}
P_{0}=\left\{ 0,0,0,0\right\} ,\qquad P_{1}=\left\{ \mu ,0,0,0\right\}
,\qquad P_{2}=\left\{ 2\mu +\alpha _{0},\alpha _{1},\alpha _{2},\alpha
_{3}\right\}  \label{a3.10}
\end{equation}%
The coordinates of vectors $\mathbf{P}_{0}\mathbf{P}_{1}$, $\mathbf{P}_{1}%
\mathbf{P}_{2}$, $\mathbf{P}_{0}\mathbf{P}_{2}$ are%
\begin{eqnarray}
\mathbf{P}_{0}\mathbf{P}_{1} &=&\left\{ \mu ,0,0,0\right\} ,\qquad \mathbf{P}%
_{1}\mathbf{P}_{2}=\left\{ \mu +\alpha _{0},\alpha _{1},\alpha _{2},\alpha
_{3}\right\} ,  \label{a3.11} \\
\mathbf{P}_{0}\mathbf{P}_{2} &=&\left\{ 2\mu +\alpha _{0},\alpha _{1},\alpha
_{2},\alpha _{3}\right\}  \label{a3.11a}
\end{eqnarray}%
Let us take into account that%
\begin{equation}
\left\vert \mathbf{P}_{0}\mathbf{P}_{1}\right\vert _{\mathrm{d}%
}^{2}=\left\vert \mathbf{P}_{0}\mathbf{P}_{1}\right\vert _{\mathrm{M}%
}^{2}=2\sigma _{\mathrm{M}}\left( P_{0},P_{1}\right) +\lambda _{0}^{2}%
\mathrm{sgn}\left( \sigma _{\mathrm{M}}\left( P_{0},P_{1}\right) \right)
\label{a3.12}
\end{equation}%
\begin{equation}
\left( \mathbf{P}_{0}\mathbf{P}_{1}.\mathbf{P}_{1}\mathbf{P}_{2}\right) _{%
\mathrm{d}}=\left( \mathbf{P}_{0}\mathbf{P}_{1}.\mathbf{P}_{1}\mathbf{P}%
_{2}\right) _{\mathrm{M}}+w\left( P_{0},P_{1},P_{1},P_{2}\right)
\label{a3.14}
\end{equation}%
Here indices "M" and "d" mean that the quantities are calculated in $%
\mathcal{G}_{\mathrm{M}}$ and $\mathcal{G}_{\mathrm{d}}$ respectively, and
for timelike vectors (\ref{a3.11})%
\begin{equation}
w\left( P_{0},P_{1},P_{1},P_{2}\right) =d\left( P_{0},P_{2}\right) +d\left(
P_{1},P_{1}\right) -d\left( P_{0},P_{1}\right) -d\left( P_{1},P_{2}\right) =-%
\frac{1}{2}\lambda _{0}^{2}  \label{a3.15}
\end{equation}%
The relation $\mathbf{P}_{0}\mathbf{P}_{1}$eqv$\mathbf{P}_{1}\mathbf{P}_{2}$
has the form of two equations
\begin{equation}
\mu \left( \mu +\alpha _{0}\right) -\frac{1}{2}\lambda _{0}^{2}=\mu
^{2}+\lambda _{0}^{2}  \label{a3.16}
\end{equation}%
\begin{equation}
\mu ^{2}=\left( \mu +\alpha _{0}\right) ^{2}-\alpha _{1}^{2}-\alpha
_{2}^{2}-\alpha _{3}^{2}  \label{a3.17}
\end{equation}%
The quantities $\alpha $ are to be determined from these equations. Solution
of equations (\ref{a3.16}), (\ref{a3.17}) has the form%
\begin{equation}
\alpha _{0}=\frac{3\lambda _{0}^{2}}{2\mu },\qquad \alpha _{1}=\lambda _{0}%
\sqrt{3+\frac{9\lambda _{0}^{2}}{4\mu ^{2}}}\sin \theta \cos \varphi ,
\label{a3.18}
\end{equation}%
\begin{equation}
\alpha _{2}=\lambda _{0}\sqrt{3+\frac{9\lambda _{0}^{2}}{4\mu ^{2}}}\sin
\theta \sin \varphi ,\qquad \alpha _{3}=\lambda _{0}\sqrt{3+\frac{9\lambda
_{0}^{2}}{4\mu ^{2}}}\cos \theta  \label{a3.19}
\end{equation}%
where the quantities $\theta $ and $\varphi $ are arbitrary.

Thus, position of the link $\mathbf{P}_{1}\mathbf{P}_{2}$ with respect to
the adjacent link $\mathbf{P}_{0}\mathbf{P}_{1}$ appears to be indefinite
(multivariant). Possible positions of the link $\mathbf{P}_{1}\mathbf{P}_{2}$
form generatrices of the cone with the axis $\mathbf{P}_{0}\mathbf{P}_{1}$%
and the angle $\phi $ at the vertex, which lies at the point $P_{1}$. The
angle $\phi $ is determined by the relation
\begin{equation}
\tan \phi =\frac{\sqrt{\alpha _{1}^{2}+\alpha _{2}^{2}+\alpha _{3}^{2}}}{\mu
+\alpha _{0}}=\frac{\lambda _{0}}{\mu \left( 1+\frac{3\lambda _{0}^{2}}{\mu
^{2}}\right) }\sqrt{3+\frac{9\lambda _{0}^{2}}{4\mu ^{2}}}\approx \frac{%
\lambda _{0}\sqrt{3}}{\mu },\qquad \text{if }\lambda _{0}\ll \mu
\label{a3.20}
\end{equation}%
If the elementary length $\lambda _{0}\rightarrow 0$, the space-time
geometry $\mathcal{G}_{\mathrm{d}}$ turns into $\mathcal{G}_{\mathrm{M}}$,
and the cone degenerates into a straight line.

Indefinite (multivariant) position of adjacent links leads to wobbling of
the world chain of the pointlike particle. Let us choose elementary length
of the space-time geometry $\mathcal{G}_{\mathrm{d}}$ in the form%
\begin{equation}
\lambda _{0}^{2}=\frac{\hbar }{bc}  \label{a3.21}
\end{equation}%
where $\hbar $ is the quantum constant, $c$ is the speed of the light and
the constant $b$ is the universal constant (\ref{a3.4}), connecting the
geometrical mass $\mu $ with the usual mass $m$. Then the statistical
description of wobbling world chains is equivalent to the quantum
description in terms of the Schr\"{o}dinger equation \cite{R91}. The quantum
constant $\hbar $ appears in the dynamics of the particle as a parameter of
the space-time geometry $\mathcal{G}_{\mathrm{d}}$. The conventional quantum
principles appear to be needless. Thus, the multivariant space-time geometry
admits one to describe quantum effects as geometric effects. Besides, the
pointlike particle mass $m$ appears to be geometrized by its connection (\ref%
{a3.4}) with the geometrical mass $\mu =\left\vert \mathbf{P}_{0}\mathbf{P}%
_{1}\right\vert _{\mathrm{d}}$.

\label{3b}We have no direct information on the space-time geometry in
microcosm. In the usual scale the space-time geometry may be considered as
continuous, because the possible discreteness of the space-time has a small
scale, which cannot be recognized in macroscopic experiments. However, in
the small scale the space-time geometry may appear to be discrete. The
discrete space-time geometry generates multivariance, which is responsible
for quantum effects. It is impossible to object anything against the
discreteness of the space-time geometry at small scale. Such a possibility
should be considered. The discrete space-time geometry is to be considered
in the framework of the physical geometry, which describes continuous and
discrete geometries, using a uniform method.\label{3e}

\section{World function of the Kaluza-Klein space-time}

The space-time geometry of Kaluza-Klein $\mathcal{G}_{\mathrm{K}}$ is given
on the 5-dimensional manifold. In the coordinate system with coordinates $%
x=\left\{ x^{0},x^{1},x^{2},x^{3},x^{5}\right\} $. Four coordinates $\left\{
x^{0},x^{1},x^{2},x^{3}\right\} =\left\{ x^{0},\mathbf{x}\right\} $ describe
position of a particle in the 4D-space-time of Minkowski, whereas the charge
coordinate $x^{5}$ describes additional characteristic of the particle,
which is responsible for interaction with the electromagnetic field.

Covariant metric tensor $\gamma _{AB}$, $A,B=0,1,2,3,5$ in the geometry $%
\mathcal{G}_{\mathrm{K}}$ is determined by the relation%
\begin{equation}
\gamma _{AB}=\left\vert \left\vert
\begin{array}{cc}
g_{ik}-a_{i}a_{k} & a_{k} \\
a_{i} & -1%
\end{array}%
\right\vert \right\vert ,\qquad i,k=0,1,2,3,\qquad A,B=0,1,2,3,5
\label{a2.1}
\end{equation}%
where $g_{ik}$, $i,k=0,1,2,3$ is the metric tensor in the conventional
4-dimensional space-time. The quantities $a_{k}$, $k=0,1,2,3$ are connected
with electromagnetic potential $A_{k}$, $k=0,1,2,3$ by means of the relation%
\begin{equation}
a_{k}=\varkappa A_{k},\qquad k=0,1,2,3  \label{a2.2}
\end{equation}%
where $\varkappa $ is some universal constant. The contravariant metric
tensor $\gamma ^{AB}$, $A,B=0,1,2,3,5$ has the form%
\begin{equation}
\gamma ^{AB}=\left\vert \left\vert
\begin{array}{cc}
g^{ik} & g^{il}a_{l} \\
g^{kl}a_{l} & -1+g^{jl}a_{j}a_{l}%
\end{array}%
\right\vert \right\vert ,\qquad i,k=0,1,2,3,\qquad A,B=0,1,2,3,5
\label{a2.3}
\end{equation}%
It is supposed that neither electromagnetic potentials $a_{k}$, nor the
metric tensor $g_{ik}$ depend on the charge coordinate $x^{5}$.

Then the action%
\begin{equation}
\mathcal{A}\left[ x\right] =\int \left\{ -m_{\mathrm{5}}c\sqrt{\gamma _{AB}%
\dot{x}^{A}\dot{x}^{B}}\right\} d\tau ,\qquad x=\left\{ x^{0}\left( \tau
\right) ,x^{1}\left( \tau \right) ,x^{2}\left( \tau \right) ,x^{3}\left(
\tau \right) ,x^{5}\left( \tau \right) \right\}  \label{a2.4}
\end{equation}%
describes the motion of a charged particle in the gravitational
field,described by the metric tensor $g_{ik}$ and in the electromagnetic
field $A_{k}$. Corresponding dynamic equations are obtained as a result of
variation of the action (\ref{a2.4}) with respect to $x^{A}$, $A=0,1,2,3,5.$%
\begin{equation}
\frac{dp_{A}}{d\tau }=-\frac{\partial }{\partial x^{A}}\left( m_{\mathrm{5}}c%
\sqrt{\gamma _{AB}\dot{x}^{A}\dot{x}^{B}}\right) ,\qquad A=0,1,2,3,5
\label{a2.5}
\end{equation}%
where%
\begin{equation}
p_{A}=-\frac{m_{\mathrm{5}}c\gamma _{AB}\dot{x}^{B}}{\sqrt{\gamma _{CD}\dot{x%
}^{C}\dot{x}^{D}}},\qquad A=0,1,2,3,5  \label{a2.6}
\end{equation}%
As far as $\gamma _{AB}$ does depend on $x^{5}$, it follows from (\ref{a2.5}%
), that the canonical momentum component $p_{5}=$const. Then, taking into
account (\ref{a2.1}), the equation (\ref{a2.7}) may be rewritten in the form%
\begin{equation}
\left( \frac{\partial S}{\partial x^{i}}+p_{5}a_{i}\right) g^{ik}\left(
\frac{\partial S}{\partial x^{k}}+p_{5}a_{k}\right) =\left( m_{\mathrm{5}%
}c\right) ^{2}+p_{5}^{2}  \label{a2.9}
\end{equation}

Comparing (\ref{a2.9}) with the Hamilton-Jacobi equation
\begin{equation}
\left( \frac{\partial S}{\partial x^{i}}+\frac{e}{c}A_{i}\right)
g^{ik}\left( \frac{\partial S}{\partial x^{k}}+\frac{e}{c}A_{k}\right)
=m^{2}c^{2}  \label{a2.10}
\end{equation}%
describing motion of a pointlike particle of mass $m$ and of charge $e$ in
4-dimensional space-time with electromagnetic potential $A_{k}$, $k=0,1,2,3$%
, one concludes that equations (\ref{a2.9}) and (\ref{a2.10}) are
equivalent, if
\begin{equation}
m=\sqrt{m_{\mathrm{5}}^{2}+c^{-2}p_{5}^{2}},\qquad p_{5}=\frac{e}{\varkappa c%
},\qquad a_{k}=\varkappa A_{k},\qquad k=0,1,2,3  \label{a2.11}
\end{equation}%
where $\varkappa $ is some universal constant.

The original action (\ref{a2.4}) has the form of the action for a geodesic
in 5-dimensional Riemannian space with the metric tensor (\ref{a2.1}). Thus,
the motion of a pointlike charged particle in the 4-dimensional Riemannian
space-time with the electromagnetic field can be described as a free motion
of a particle in the 5-dimensional Riemannian space-time. The electric
charge $e$ of the particle is geometrized in the sense, that it appears to
be connected with the component $p_{5}$ of the particle momentum along the
fifth (charge) coordinate $x^{5}$.

However, the fifth coordinate $x^{5}$ is unobservable, and one tries to
explain this circumstance by the hypothesis, that the space-time of
Kaluza-Klein is thin in the direction of the fifth coordinate $x^{5}$. One
supposes, that the space-time of Kaluza-Klein is compactified in the
direction of fifth coordinate $x^{5}$, i.e. the points with coordinates $%
\left\{ x^{0},x^{1},x^{2},x^{3},x^{5}\right\} $ and $\left\{
x^{0},x^{1},x^{2},x^{3},x^{5}+2kL\right\} $ coincide, where $L$ is some
universal constant and $k$ is any integer number.

\section{Discrimination properties of the Kaluza-Klein geometry
compactification}

We shall try to analyze influence of compactification on the Kaluza-Klein
geometry $\mathcal{G}_{\mathrm{K}}$. For simplicity we shall consider the
case, when the gravitational field and the electromagnetic one are absent.
Then the metric tensor (\ref{a2.1}) takes the form $\gamma _{AB}=$diag$%
\left( c^{2},-1,-1,-1,-1\right) $ and $a_{k}=0$, $k=0,1,2,3$. Geodesics $%
\mathcal{L}_{P_{0}P_{1}}$, passing through points $P_{0}$ and $P_{1}$ with
coordinates
\begin{equation}
P_{0}=\left\{ 0,0,0,0,0\right\} ,\qquad P_{1}=\left\{
y^{0},y^{1},y^{2},y^{3},y^{5}\right\} ,\qquad
y^{0},y^{1},y^{2},y^{3},y^{5}\in \mathbb{R}  \label{a2.12}
\end{equation}%
have the form%
\begin{equation}
x^{k}=y^{k}\tau ,\qquad x^{5}=\left( y^{5}+2nL\right) \tau ,\qquad k=0,1,2,3
\label{a2.14}
\end{equation}%
where $\tau $ is a parameter along the geodesic, and $n$ is an arbitrary
integer number. The compactification may be considered as a conglutination
of points with coordinates $\left\{ x^{0},x^{1},x^{2},x^{3},x^{5}-L\right\} $
and $\left\{ x^{0},x^{1},x^{2},x^{3},x^{5}+L\right\} $. As a result one
obtains a "cylinder" instead of a plane. The compactification distinguishes
the space-time direction of the coordinate $x^{5}$ in the sense that it
forbids space-time rotations, including the coordinate $x^{5}$.

Defining the world function $\sigma _{\mathrm{K}}\left( P_{0},P_{1}\right) $
by means of (\ref{a1.7}) as an integral along the geodesic, connecting
points $P_{0}$ and $P_{1}$, one obtains a many-valued world function,
because there are many geodesics of different length, connecting the points $%
P_{0}$ and $P_{1}$. If the space-time geometry is constructed according to
conventional method on the basis of the linear vector space, the metric
tensor is a primary quantity, whereas the world function is a secondary
(derivative) quantity. In this case one may accept situation with
many-valued world function, and one may try to interpret this fact in some
way.

However, if the space-time geometry is a physical geometry, where the world
function is the primary fundamental quantity, one cannot accept a
many-valued primary quantity. One needs to use a single-valued world
function and to choose only one of many possible variants of the geodesic (%
\ref{a2.14}). One obtains different space-time geometries for different
choice of the geodesic (\ref{a2.14}), determining the world function.

\label{0b}The single-valued world function restricts possible values of
electric charge, considered as a momentum along the fifth coordinate $x^{5}$
in the space-time of Kaluza-Klein. As a result of the single-valued world
function the electric charge of an elementary particle appears to be
restricted. Compactification with many-valued world function does not need
such a restriction. \label{0e}

We consider the simplest case, when the world function is defined as
integral (\ref{a1.7}) along the "shortest" geodesic, corresponding to the
geodesic (\ref{a2.14}). This geodesic makes less, than one convolution
around the "cylinder". In this case the world function depends on the
standartized value $x_{\mathrm{st}}^{5}$ of the coordinate $x^{5}$%
\begin{equation}
\sigma _{\mathrm{K}}\left( x,x^{\prime }\right) =\frac{1}{2}\left( \left(
x^{0}-x^{\prime 0}\right) ^{2}-\left( \mathbf{x}-\mathbf{x}^{\prime }\right)
^{2}-\left( \left( x^{5}-x^{\prime 5}\right) _{\mathrm{st}}\right)
^{2}\right)  \label{a2.15}
\end{equation}%
where $\mathbf{x}=\left\{ x^{1},x^{2},x^{3}\right\} $
\begin{equation}
x_{\mathrm{st}}=\left\{
\begin{array}{lll}
2L\left\{ \frac{x}{2L}\right\} & \text{if} & 2L\left\{ \frac{x}{2L}\right\}
\leq L \\
2L\left\{ \frac{x}{2L}\right\} -2L & \text{if} & L<2L\left\{ \frac{x}{2L}%
\right\}%
\end{array}%
\right. ,\qquad 2L\left\{ \frac{x}{2L}\right\} \in \lbrack 0,2L)
\label{a2.16}
\end{equation}%
Here $\left\{ x\right\} $ means the fractional part of a decimal number $x$,
and $\left[ x\right] $ is the integer part of $x$. In other words, $\left[ x%
\right] $ and $\left\{ x\right\} $ are defined by relations.

\begin{equation}
\left[ x\right] =\max \left( k\in \mathbb{Z}|k\leq x\right) ,  \label{a2.17}
\end{equation}%
where $\mathbb{Z}$ is the set of all integer numbers.%
\begin{equation}
\left\{ x\right\} =x-\left[ x\right]  \label{a2.18}
\end{equation}%
The coordinate $x_{\mathrm{st}}^{5}$ is a standartized coordinate $x_{%
\mathrm{st}}^{5}\in (-L,L]$, although formally $x^{5}\in \mathbb{R}$. The
expression $\left( x^{5}-x^{\prime 5}\right) _{\mathrm{st}}\in (-L,L]$,
although formally $x^{5},x^{\prime 5}\in \mathbb{R}$. We have
\begin{equation}
\left( x_{\mathrm{st}}-x_{\mathrm{st}}^{\prime }\right) _{\mathrm{st}%
}=\left\{
\begin{array}{lll}
x_{\mathrm{st}}-x_{\mathrm{st}}^{\prime } & \text{if} & -L<x_{\mathrm{st}%
}-x_{\mathrm{st}}^{\prime }\leq L \\
-2L+x_{\mathrm{st}}-x_{\mathrm{st}}^{\prime } & \text{if} & L<x_{\mathrm{st}%
}-x_{\mathrm{st}}^{\prime }\leq 3L \\
2L+x_{\mathrm{st}}-x_{\mathrm{st}}^{\prime } & \text{if} & -3L<x_{\mathrm{st}%
}-x_{\mathrm{st}}^{\prime }\leq -L%
\end{array}%
\right.  \label{a2.19}
\end{equation}%
The choice of the world function $\sigma _{\mathrm{K}}$ in the form (\ref%
{a2.15}), (\ref{a2.16}) corresponds to the geodesic (\ref{a2.14}), which
makes less, than one convolution around the "cylinder". The world function (%
\ref{a2.15}), (\ref{a2.16}) is zero-variant with respect to some vectors.

Let us consider the points of two adjacent vectors of a world chain.%
\begin{eqnarray}
P_{0} &=&\left\{ 0,0,0,0,0\right\} ,\qquad P_{1}=\left\{
s_{0},s_{1},s_{2},s_{3},l\right\} ,  \label{a2.20} \\
P_{2} &=&\left\{ 2s_{0}+\alpha _{0},2s_{1}+\alpha _{1},2s_{2}+\alpha
_{2},2s_{3}+\alpha _{3},2l+\alpha _{5}\right\}  \label{a2.21c}
\end{eqnarray}%
\begin{eqnarray}
\mathbf{P}_{0}\mathbf{P}_{1} &=&s=\left\{ s_{0},s_{1},s_{2},s_{3},l\right\} ,
\label{a2.21a} \\
\mathbf{P}_{1}\mathbf{P}_{2} &=&s+\alpha =\left\{ s_{0}+\alpha
_{0},s_{1}+\alpha _{1},s_{2}+\alpha _{2},s_{3}+\alpha _{3},l+\alpha
_{5}\right\}  \label{a2.21d}
\end{eqnarray}%
\begin{equation}
\mathbf{P}_{0}\mathbf{P}_{2}=2s+\alpha =\left\{ 2s_{0}+\alpha
_{0},2s_{1}+\alpha _{1},2s_{2}+\alpha _{2},2s_{3}+\alpha _{3},2l+\alpha
_{5}\right\}  \label{a2.21b}
\end{equation}%
We shall show, that if the fifth coordinate $x^{5}=l$ satisfies the relation%
\begin{equation}
\left\vert l\right\vert >\frac{L}{2}  \label{a2.22}
\end{equation}%
then the vector $\mathbf{P}_{1}\mathbf{P}_{2}$, which is equivalent to
vector $\mathbf{P}_{0}\mathbf{P}_{1}$ does not exist. It means, that the
world chain of a free pointlike particle with the link $\mathbf{P}_{0}%
\mathbf{P}_{1}$ cannot exist.

The equivalence conditions $\mathbf{P}_{0}\mathbf{P}_{1}$eqv$\mathbf{P}_{1}%
\mathbf{P}_{2}$ for vectors (\ref{a2.21a}), (\ref{a2.21d}) are written in
the form
\begin{equation}
\left\vert \mathbf{P}_{0}\mathbf{P}_{1}\right\vert _{\mathrm{K}%
}^{2}=\left\vert \mathbf{P}_{1}\mathbf{P}_{2}\right\vert _{\mathrm{K}}^{2}
\label{a2.23}
\end{equation}%
\begin{equation}
\left( \mathbf{P}_{0}\mathbf{P}_{1}.\mathbf{P}_{1}\mathbf{P}_{2}\right) _{%
\mathrm{K}}=\left\vert \mathbf{P}_{0}\mathbf{P}_{1}\right\vert _{\mathrm{K}%
}^{2}  \label{a2.24}
\end{equation}%
where index "K" means, that the corresponding quantities are taken in the
geometry (\ref{a2.15}).

We suppose, that the vector $\mathbf{P}_{0}\mathbf{P}_{1}$ is timelike in
the sense, that%
\begin{equation}
s_{0}^{2}>L^{2}+\mathbf{s}^{2},\qquad \mathbf{s}=\left\{
s_{1},s_{2},s_{3}\right\}  \label{a2.25}
\end{equation}%
As far as
\begin{equation}
\left( \mathbf{P}_{0}\mathbf{P}_{1}.\mathbf{P}_{1}\mathbf{P}_{2}\right) _{%
\mathrm{K}}=\sigma _{\mathrm{K}}\left( P_{0},P_{2}\right) -\sigma _{\mathrm{K%
}}\left( P_{0},P_{1}\right) -\sigma _{\mathrm{K}}\left( P_{1},P_{2}\right)
\label{a2.26}
\end{equation}%
the equations (\ref{a2.23}), (\ref{a2.24}) are written in the form%
\begin{equation}
s_{0}^{2}-\mathbf{s}^{2}-l^{2}=\left( s_{0}+\alpha _{0}\right) ^{2}-\left(
\mathbf{s+\alpha }\right) ^{2}-\left( \left( l+\alpha _{5}\right) _{\mathrm{%
st}}\right) ^{2}  \label{a2.27}
\end{equation}%
\begin{equation}
\left( 2s_{0}+\alpha _{0}\right) ^{2}-\left( 2\mathbf{s+\alpha }\right)
^{2}-\left( 2l+\alpha _{5}\right) _{\mathrm{st}}^{2}=4\left( s_{0}^{2}-%
\mathbf{s}^{2}-l^{2}\right)  \label{a2.28}
\end{equation}

Taking sum of equations (\ref{a2.28}) and (\ref{a2.27}), one obtains%
\begin{equation}
2s_{0}\alpha _{0}-2\mathbf{s\alpha }-\left( 2l+\alpha _{5}\right) _{\mathrm{%
st}}^{2}+\left( l+\alpha _{5}\right) _{\mathrm{st}}^{2}=-3l^{2}
\label{a2.29}
\end{equation}%
\begin{equation}
\alpha _{0}=\frac{2\mathbf{s\alpha +}\left( 2l+\alpha _{5}\right) _{\mathrm{%
st}}^{2}-\left( l+\alpha _{5}\right) _{\mathrm{st}}^{2}-3l^{2}}{2s_{0}}
\label{a2.30}
\end{equation}

Substituting (\ref{a2.30}) in (\ref{a2.27}), one obtains%
\begin{equation}
\mathbf{\alpha }^{2}=\left( 2l+\alpha _{5}\right) _{\mathrm{st}}^{2}-2\left(
l+\alpha _{5}\right) _{\mathrm{st}}^{2}-2l^{2}+\left( \frac{2\mathbf{s\alpha
+}\left( 2l+\alpha _{5}\right) _{\mathrm{st}}^{2}-\left( l+\alpha
_{5}\right) _{\mathrm{st}}^{2}-3l^{2}}{2s_{0}}\right) ^{2}  \label{a2.31}
\end{equation}

Let us set%
\begin{equation}
\beta =\beta _{\mathrm{st}}=\left( l+\alpha _{5}\right) _{\mathrm{st}}
\label{a2.32}
\end{equation}%
Then%
\begin{equation}
\left( 2l+\alpha _{5}\right) _{\mathrm{st}}=\left( l+\beta \right) _{\mathrm{%
st}}=l+\beta +\gamma  \label{a2.33}
\end{equation}%
where%
\begin{equation}
\gamma =\left\{
\begin{array}{ccc}
0 & \text{if} & -L<l+\beta \leq L \\
-2L & \text{if} & L<l+\beta \leq 3L \\
2L & \text{if} & -3L<l+\beta \leq -L%
\end{array}%
\right. =\left\{
\begin{array}{ccc}
0 & \text{if} & -L-l<\beta \leq L-l \\
-2L & \text{if} & L-l<\beta \leq 3L-l \\
2L & \text{if} & -3L-l<\beta \leq -L-l%
\end{array}%
\right.  \label{a2.35}
\end{equation}

Note, that we are interested in the quantity $\beta $, because it is the
fifth coordinate of the vector $\mathbf{P}_{1}\mathbf{P}_{2}$, which is
determined to within $2kL$, where $k$ is an arbitrary integer number%
\begin{eqnarray}
\mathbf{P}_{1}\mathbf{P}_{2} &=&\left\{ s_{0}+\alpha _{0},s_{1}+\alpha
_{1},s_{2}+\alpha _{2},s_{3}+\alpha _{3},\beta \right\}  \notag \\
&=&\left\{ s_{0}+\alpha _{0},s_{1}+\alpha _{1},s_{2}+\alpha
_{2},s_{3}+\alpha _{3},\beta +2kL\right\}  \label{a2.35b}
\end{eqnarray}

Substituting (\ref{a2.32}) and (\ref{a2.33}) in (\ref{a2.31}), one obtains
after transformations%
\begin{eqnarray}
\mathbf{\alpha }^{2} &=&+\frac{2\mathbf{s\alpha }l\left( \beta -l\right) }{%
s_{0}^{2}}+\frac{\left( \mathbf{s\alpha }\right) ^{2}}{s_{0}^{2}}-\left(
l-\beta \right) ^{2}\left( 1-\frac{l^{2}}{s_{0}^{2}}\right) +\gamma
^{2}\left( \frac{\frac{1}{2}\gamma +\left( l+\beta \right) }{s_{0}}\right)
^{2}  \notag \\
&&+\gamma \left( \gamma +2\left( l+\beta \right) \right) \left( 1+\frac{%
l\left( \beta -l\right) +\mathbf{s\alpha }}{s_{0}^{2}}\right)  \label{b2.36a}
\end{eqnarray}

Let us consider the case, when%
\begin{equation}
\gamma =0,\qquad -L<\beta +l\leq L,  \label{b2.37}
\end{equation}
Then one obtains from (\ref{b2.36a})%
\begin{equation}
\mathbf{\alpha }^{2}=-\left( l-\beta \right) ^{2}\left( 1-\frac{l^{2}}{%
s_{0}^{2}}\right) +\left( \frac{\mathbf{s\alpha }}{s_{0}}\right) ^{2}-2\frac{%
\mathbf{s\alpha }}{s_{0}^{2}}l\left( l-\beta \right)  \label{b2.38}
\end{equation}

One can see, that the equation (\ref{b2.38}) has the evident solution%
\begin{equation}
\beta =l,\qquad \mathbf{\alpha }=\left( \alpha _{1},\alpha _{2},\alpha
_{3}\right) =\left( 0,0,0\right) ,\qquad \alpha _{0}=0  \label{b2.38a}
\end{equation}%
. It follows from (\ref{b2.37}) and (\ref{b2.38a}), that%
\begin{equation}
-L/2<l\leq L/2,\qquad -L/2<\beta \leq L/2  \label{b2.38b}
\end{equation}

To obtain other solutions, let us set%
\begin{equation}
\beta =l+\varepsilon  \label{b2.39}
\end{equation}%
One obtains instead of (\ref{b2.38})%
\begin{equation}
\mathbf{\alpha }^{2}=-\varepsilon ^{2}\left( 1-\frac{l^{2}}{s_{0}^{2}}%
\right) +\left( \frac{\mathbf{s\alpha }}{s_{0}}\right) ^{2}+2\frac{\mathbf{%
s\alpha }}{s_{0}^{2}}l\varepsilon  \label{b2.40}
\end{equation}%
Or%
\begin{equation}
\dsum\limits_{\beta }\left( s_{0}^{2}\alpha _{\beta }^{2}-2ls_{\beta
}\varepsilon \alpha _{\beta }-\dsum\limits_{\nu }s_{\beta }s_{\nu }\alpha
_{\beta }\alpha _{\nu }\right) +\varepsilon ^{2}\left(
s_{0}^{2}-l^{2}\right) =0  \label{b2.41}
\end{equation}%
We are to find such spacelike vectors $\left\{ \alpha _{0},\alpha
_{1},\alpha _{2},\alpha _{3},\varepsilon \right\} $ and such a value of the
variable $l$, which satisfy the equation (\ref{b2.41}).

Let us choose the axis $x^{1}$ along the 3-vector $\mathbf{s}$. Equation (%
\ref{b2.41}) takes the form%
\begin{equation}
\left( s_{0}^{2}-s_{1}^{2}\right) \alpha _{1}^{2}+s_{0}^{2}\alpha
_{2}^{2}+s_{0}^{2}\alpha _{3}^{2}+\left( s_{0}^{2}-l^{2}\right) \varepsilon
^{2}-2ls_{1}\varepsilon \alpha _{1}=0  \label{b2.41a}
\end{equation}%
Lhs of equation (\ref{b2.41a}) is a quadratic form with respect to variables
$\left\{ \alpha _{1},\alpha _{2},\alpha _{3},\varepsilon \right\} .$The
matrix of the quadratic form of the equation (\ref{b2.41a}) has the form%
\begin{equation}
\left\vert \left\vert
\begin{array}{cccc}
s_{0}^{2}-s_{1}^{2} & 0 & 0 & -ls_{1} \\
0 & s_{0}^{2} & 0 & 0 \\
0 & 0 & s_{0}^{2} & 0 \\
-ls_{1} & 0 & 0 & \left( s_{0}^{2}-l^{2}\right)%
\end{array}%
\right\vert \right\vert  \label{b2.42}
\end{equation}%
Eigenvectors and eigenvalues of the quadratic form (\ref{b2.42}) have the
form,
\begin{equation}
\left\{
\begin{array}{c}
0 \\
1 \\
0 \\
0%
\end{array}%
,%
\begin{array}{c}
0 \\
0 \\
1 \\
0%
\end{array}%
,%
\begin{array}{c}
-\frac{l}{s_{1}} \\
0 \\
0 \\
1%
\end{array}%
\right\} \leftrightarrow s_{0}^{2},\left\{
\begin{array}{c}
\frac{1}{l}s_{1} \\
0 \\
0 \\
1%
\end{array}%
\right\} \leftrightarrow -l^{2}+s_{0}^{2}-s_{1}^{2}  \label{b2.43}
\end{equation}

The equation (\ref{b2.41}) has trivial solution (\ref{b2.38a}): $\mathbf{%
\alpha =}\left\{ 0,0,0\right\} ,\varepsilon =0$. The equation (\ref{b2.41})
has nontrivial solution if, at least, one of eigenvalues of the matrix (\ref%
{b2.42}) vanishes. For timelike vector $\left\{ s_{0},\mathbf{s,}l\right\} $
we have
\begin{equation}
s_{0}^{2}>\mathbf{s}^{2}+l^{2}  \label{b2.44}
\end{equation}%
Let us try to find such values of the variable $l$, for which the eigenvalue
vanishes. The first eigenvalue of (\ref{b2.43}) is positive always. The
second eigenvalue of (\ref{b2.43}) vanishes, if
\begin{equation}
s_{0}^{2}-s_{1}^{2}-l^{2}=0  \label{b2.45}
\end{equation}%
Fulfilment of equation (\ref{b2.45}) is impossible, because of (\ref{b2.44}%
). It means, that eigenvalues of the matrix (\ref{b2.42}) do not vanish and
the equation (\ref{b2.41}) has only trivial solution
\begin{equation}
\mathbf{\alpha =}\left\{ 0,0,0\right\} ,\quad \alpha _{0}=0,\quad
\varepsilon =0,\quad \alpha _{5}=0,\quad \beta =l\qquad \mathrm{if}%
-L/2<l\leq L/2  \label{b2.47}
\end{equation}

Let us consider the case

\begin{equation}
\gamma =-2L,\qquad L<l+\beta \leq 3L  \label{b2.47a}
\end{equation}%
In the nonrelativistic case $\mathbf{s}^{2},l^{2},L^{2}\ll s_{0}^{2}$ the
equation (\ref{b2.36a}) takes the form%
\begin{equation*}
\mathbf{\alpha }^{2}=\left( l+\beta +\gamma \right) ^{2}-2\beta ^{2}-2l^{2}
\end{equation*}%
\begin{equation}
\mathbf{\alpha }^{2}+\left( l-\beta \right) ^{2}-2L\left( 2L-2\left( l+\beta
\right) \right) =0  \label{b2.47b}
\end{equation}%
This equation can be written in the form%
\begin{equation}
l+\beta =-\frac{\mathbf{\alpha }^{2}+\left( l-\beta \right) ^{2}}{4L}+L\leq L
\label{b2.47d}
\end{equation}%
As it follows from comparison of (\ref{b2.47a}) and of (\ref{b2.47d}) the
equation (\ref{b2.47b}) has no solution, satisfying the inequality (\ref%
{b2.47a}) even in the case when
\begin{equation}
\mathbf{\alpha }=\left\{ 0,0,0\right\} ,\qquad l=\beta =L/2  \label{b2.42a}
\end{equation}

Let us consider the case
\begin{equation}
\gamma =2L,\qquad -3L<l+\beta \leq -L  \label{b2.42b}
\end{equation}%
In the nonrelativistic case $\mathbf{s}^{2},l^{2},L^{2}\ll s_{0}^{2}$ the
equation (\ref{b2.36a}) takes the form%
\begin{equation}
\mathbf{\alpha }^{2}=-\left( l-\beta \right) ^{2}+2L\left( 2L+2\left(
l+\beta \right) \right)  \label{b2.42c}
\end{equation}%
According with (\ref{b2.42b}) this equation can be written in the form%
\begin{equation}
l+\beta =\frac{\mathbf{\alpha }^{2}+\left( l-\beta \right) ^{2}}{4L}-L\leq -L
\label{b2.42d}
\end{equation}%
As it follows from the equation (\ref{b2.42d}) and inequality (\ref{b2.42b}%
), that the solution of equation (\ref{b2.42d}) has the form%
\begin{equation}
\mathbf{\alpha }=\left\{ 0,0,0\right\} ,\qquad l=\beta =-L/2  \label{a2.39}
\end{equation}%
Uniting (\ref{b2.47}) with (\ref{a2.39}), one obtains, that if vectors $%
\mathbf{P}_{0}\mathbf{P}_{1}$ and $\mathbf{P}_{1}\mathbf{P}_{2}$ are
timelike in the sense (\ref{a2.25}), the unique solution of (\ref{a2.31}) is
\begin{equation}
\mathbf{\alpha }=0,\qquad \alpha _{5}=0,\qquad \alpha _{0}=0,\qquad -\frac{L%
}{2}\leq l\leq \frac{L}{2}  \label{a2.40}
\end{equation}

Thus, one obtains, that at the point $P_{1}$ there is only one vector $%
\mathbf{P}_{1}\mathbf{P}_{2}=\left\{ s_{0},s_{1},s_{2},s_{3},l\right\} $,
which is equivalent to the vector $\mathbf{P}_{0}\mathbf{P}_{1}=\left\{
s_{0},s_{1},s_{2},s_{3},l\right\} $ at the point $P_{0}$. This equivalence
takes place only, if $l$ satisfies the relation
\begin{equation}
\left\vert l\right\vert \leq \frac{L}{2}  \label{a2.44}
\end{equation}%
If the relation (\ref{a2.44}) is not satisfied, at the point $P_{1}$ there
is no vector $\mathbf{P}_{1}\mathbf{P}_{2}$, which is equivalent to the
vector $\mathbf{P}_{0}\mathbf{P}_{1}=\left\{
s_{0},s_{1},s_{2},s_{3},l\right\} $.

If the Kaluza-Klein geometry is not compactificated the vectors $\mathbf{P}%
_{0}\mathbf{P}_{1}=\left\{ s_{0},s_{1},s_{2},s_{3},l\right\} $ and $\mathbf{P%
}_{1}\mathbf{P}_{2}=\left\{ s_{0},s_{1},s_{2},s_{3},l\right\} $ are
equivalent at any value of the charge $l$ (at $l^{2}<s_{0}^{2}-\mathbf{s}%
^{2} $). Thus, the compactification discriminates large values of the charge
coordinate $x^{5}=l$. Influence of compactification reminds influence of a
potential hole with infinite high walls placed at the values $-L/2$ and $L/2$
of the fifth coordinate $x^{5}$. In both cases displacement of a particle in
the fifth direction is restricted. In the case of the potential hole only
displacement (but not momentum $p_{5})$ is restricted. In the case of
compactification the value of momentum $p_{5}$ (electric charge) is
restricted, when the physical space-time geometry is used. In this case the
links of the world chain have finite length and a discrimination of large
values of the electric charge appears. At the conventional approach to the
Kaluza-Klein geometry, based on the linear vector space, the
compactification does not discriminate any values of the electric charge (in
the case of classical dynamics), because the length of the particle world
chain links is considered to be infinitesimal.\label{001}

The charge coordinate $x^{5}=l$ describes a displacement of the particle in
the fifth direction $x^{5}$. In the physical geometry the component $x^{5}=l$
of the vector $\mathbf{P}_{0}\mathbf{P}_{1}=\left\{ s_{0},0,0,0,l\right\} $,
is simultaneously a component $p_{5}$ of the momentum vector (the electric
charge to within a factor). The discrimination of values of the quantity $l$
is a discrimination of the charge component $p_{5}$ of the momentum vector $%
\mathbf{P}_{0}\mathbf{P}_{1}$, i.e. it is a discrimination of the particle
electric charge. \label{1b} Conventional method of compactification, using a
single-valued metric tensor (but a many-valued world function), does not
admit one to obtain a restriction of the module of the electric charge of an
elementary particle. It generates only periodical dependence of the particle
state on the fifth coordinate. Of course, this periodical dependence relates
only to the state of the statistical ensemble, but not to the state of a
single particle. Instead of the dynamic equations for a single free particle
\begin{equation}
\frac{d}{dt}x^{A}\left( t\right) =v^{A}\left( t\right) \mathbf{,\qquad }%
\frac{dv^{A}\left( t\right) }{dt}=0,\qquad A=1,2,3,5  \label{a2.44a}
\end{equation}%
one should consider dynamic equations for the statistical ensemble
consisting of free particles, whose motion is described by equations (\ref%
{a2.44a}). In particular, if the state of this ensemble is described by the
wave function $\psi \left( t,\mathbf{x},x^{5}\right) $, the wave function is
to be a periodical function of the fifth coordinate $x^{5}$

\begin{equation}
\psi \left( t,\mathbf{x,}x^{5}\right) =\psi \left( t,\mathbf{x,}%
x^{5}+2kL\right)  \label{a2.44c}
\end{equation}%
where $k$ is any integer number. In the framework of quantum mechanics this
periodicity leads to that result, that the operator of the electric charge $%
-i\hbar \partial /\partial x^{5}$ has eigenvalues which are multiple to some
elementary electric charge.

After compactification the single-valued world function restricts the
particle displacement in the direction of fifth coordinate. However, in
general, it does not restrict the charge component $p_{5}$ of the momentum
vector. The the charge component $p_{5}$ is restricted, if (1) \textit{the
links of the world chain have a finite length} and (2) \textit{the world
function is single-valued. }If one of these conditions is violated, the
value of the charge component $p_{5}$ of the momentum vector may be not
restricted.

In particular, at the conventional approach, when the world function becomes
to be many-valued after compactification, the particle displacement along
the direction $x^{5}$, and momentum $p_{5}$ of a particle remain to be
unrestricted. If the world function is single-valued after compactification,
but the world chain links are infinitesimal, the particle displacement along
the direction $x^{5}$ appears to be restricted, but the momentum $p_{5}$
remains to be unrestricted. In particular, in the discrete space-time
geometry, where the links of the world chain cannot be infinitesimal, the
momentum $p_{5}$ appears to be restricted, if the world function is made
single-valued after compactification.

It is well known, that stable elementary particles have the electric charge $%
0,\pm e_{0}$, where $e_{0}$ is the elementary charge. Only short-living
resonances have multiple charges. Apparently, they are bound states of
several elementary particles. As to quarks, which have fractional electric
charge, they cannot be extracted from stable elementary particles. Quarks
are rather elements of a structure of elementary particles, than elementary
particles themselves.

Thus, experimental data confirm a reasonable supposition on
single-valuedness of the world function after compactification.\label{1e}

\section{Compactification in the discrete Kaluza-Klein space-time}

Let us consider compactification of the discrete Kaluza-Klein space-time.
The world function has the form%
\begin{equation}
\sigma _{\mathrm{dK}}\left( x,x^{\prime }\right) =\sigma _{\mathrm{K}}\left(
x,x^{\prime }\right) +\frac{\lambda _{0}^{2}}{2}\text{sgn}\left( \sigma _{%
\mathrm{K}}\left( x,x^{\prime }\right) \right)   \label{a5.1}
\end{equation}%
where $\sigma _{\mathrm{K}}$ is determined by the relation (\ref{a2.15}), (%
\ref{a2.16}). As far as the space-time geometry with the world function (\ref%
{a5.1}) is discrete, the world chains of particles have links of a finite
length, because in the discrete space-time geometry the link length cannot
be infinitesimal. We consider two timelike vectors $\mathbf{P}_{0}\mathbf{P}%
_{1}$ and $\mathbf{P}_{1}\mathbf{P}_{2}$ of the world chain. The vectors are
determined by the relations (\ref{a2.20}) - (\ref{a2.21b}). These vectors
are supposed to be equivalent and to satisfy the relations of the type (\ref%
{a2.23}), (\ref{a2.24}).

The equations are rewritten in the developed form
\begin{equation}
s_{0}^{2}-\mathbf{s}^{2}-l^{2}=\left( s_{0}+\alpha _{0}\right) ^{2}-\left(
\mathbf{s+\alpha }\right) ^{2}-\left( l+\alpha _{5}\right) _{\mathrm{st}}^{2}
\label{a5.4}
\end{equation}%
\begin{equation}
\left( \left( 2s_{0}+\alpha _{0}\right) ^{2}-\left( 2\mathbf{s+\alpha }%
\right) ^{2}-\left( 2l+\alpha _{5}\right) _{\mathrm{st}}^{2}\right) +\lambda
_{0}^{2}=4\left( \left( s_{0}^{2}-\mathbf{s}^{2}-l^{2}\right) +\lambda
_{0}^{2}\right)  \label{a5.5}
\end{equation}%
Combining equations (\ref{a5.4}) (\ref{a5.5}), one obtains
\begin{equation}
\alpha _{0}=\frac{2\mathbf{s\alpha +}\left( 2l+\alpha _{5}\right) _{\mathrm{%
st}}^{2}-\left( l+\alpha _{5}\right) _{\mathrm{st}}^{2}-3l^{2}+3\lambda
_{0}^{2}}{2s_{0}}  \label{a5.7}
\end{equation}

Substituting (\ref{a5.7}) in (\ref{a5.4}), one obtains%
\begin{equation}
\mathbf{\alpha }^{2}=\left( 2l+\alpha _{5}\right) _{\mathrm{st}}^{2}-2\left(
l+\alpha _{5}\right) _{\mathrm{st}}^{2}-2l^{2}+3\lambda _{0}^{2}+\left(
\frac{2\mathbf{s\alpha +}\left( 2l+\alpha _{5}\right) _{\mathrm{st}%
}^{2}-\left( l+\alpha _{5}\right) _{\mathrm{st}}^{2}-3l^{2}+3\lambda _{0}^{2}%
}{2s_{0}}\right) ^{2}  \label{a5.8}
\end{equation}

Or%
\begin{equation}
\mathbf{\alpha }^{2}=\left( l+\gamma \right) \left( l+2\beta +\gamma \right)
-\beta ^{2}-2l^{2}+3\lambda _{0}^{2}+\left( \frac{2\mathbf{s\alpha +}\left(
l+\gamma \right) \left( l+2\beta +\gamma \right) -3l^{2}+3\lambda _{0}^{2}}{%
2s_{0}}\right) ^{2}  \label{a5.9}
\end{equation}%
where $\gamma $ is determined by (\ref{a2.35})

We shall consider only nonrelativistic case $\mathbf{s}^{2},l^{2},L^{2}\ll
s_{0}^{2}$ for timelike vectors $\mathbf{P}_{0}\mathbf{P}_{1}$ and $\mathbf{P%
}_{1}\mathbf{P}_{2}$. In the case, when $\gamma =0$ and according to (\ref%
{a2.32}), (\ref{a2.35})%
\begin{equation}
-L<l+\beta \leq L  \label{a5.10}
\end{equation}%
one obtains from (\ref{a5.9})
\begin{equation}
\mathbf{\alpha }^{2}+\left( \beta -l\right) ^{2}=r_{1}^{2},\qquad
r_{1}^{2}=3\lambda _{0}^{2}  \label{a5.11}
\end{equation}%
The relation (\ref{a5.11}) describes a sphere of the radius
\begin{equation}
r_{1}=\sqrt{3}\lambda _{0}  \label{a5.12}
\end{equation}%
with the center $\left\{ \mathbf{\alpha }_{\mathrm{c}},\beta _{\mathrm{c}%
}\right\} =\left\{ \mathbf{0},l\right\} $ in the 4-dimensional space of
coordinates $\left\{ \mathbf{\alpha },\beta \right\} =\left\{ \alpha
_{1},\alpha _{2},\alpha _{3},\beta \right\} $.

Solution of equations (\ref{a5.11}) has the form%
\begin{eqnarray}
\alpha _{1} &=&r_{1}\sin \theta \sin \phi _{2}\sin \phi _{3},\qquad \alpha
_{2}=r_{1}\sin \theta \sin \phi _{2}\cos \phi _{3},  \label{a5.14} \\
\alpha _{3} &=&r_{1}\sin \theta \cos \phi _{2},\qquad \beta =l+r_{1}\cos
\theta ,\qquad \alpha _{0}=0,\qquad  \label{a5.15}
\end{eqnarray}%
which is valid for$\mathrm{\ }$%
\begin{equation}
r_{1}\cos \theta \leq L-2l.  \label{a5.15a}
\end{equation}%
Here $r_{1}$ is defined by relation (\ref{a5.12}), and $\theta ,\phi
_{2},\phi _{3}$ are arbitrary numbers. Although the solution (\ref{a5.14}), (%
\ref{a5.15}) is many-valued, but it is placed at the distance of the order
of $\lambda _{0}$ from the single-valued solution (\ref{a2.40}). At $\lambda
_{0}=0$ the radius $r_{1}$ of the sphere (\ref{a5.11}) vanishes and the
solution (\ref{a5.14}), (\ref{a5.15}) coincides with (\ref{a2.40}).

We are interested in behaviour of the solutions the equation (\ref{a5.11})
near the boundary $x^{5}=L/2$. The sphere (\ref{a5.11}) lies either
completely inside the region $-L/2<x^{5}<L/2$, or intersects the boundary $%
x^{5}=L/2$. In the last case we use the designation
\begin{equation}
l=\frac{L}{2}+\delta ,\qquad -\frac{r_{1}}{2}<\delta \leq \frac{r_{1}}{2}
\label{a5.20a}
\end{equation}%
then according to (\ref{a5.15})%
\begin{equation}
\beta =\frac{L}{2}+\delta +r_{1}\cos \theta  \label{b5.21}
\end{equation}%
The angle $\theta _{\mathrm{\max }}$ of intersection of the sphere (\ref%
{a5.11}) with the boundary $x^{5}=L/2$ is defined by the relation
\begin{equation}
\cos \theta _{\mathrm{\max }}=-\frac{2\delta }{r_{1}}  \label{b5.22a}
\end{equation}%
which follows from (\ref{a5.15a}). Position of the sphere section is
determined by%
\begin{equation}
\beta _{\mathrm{\max }}=L/2+\delta +r_{1}\cos \theta _{\mathrm{\max }%
}=L/2-\delta  \label{b5.22c}
\end{equation}%
which corresponds to the condition%
\begin{equation}
\beta _{\mathrm{\max }}+l=L  \label{b5.22d}
\end{equation}%
Radius $R$ of the sphere section has the form%
\begin{equation}
R=r_{1}\sin \theta _{\mathrm{\max }}=r_{1}\sqrt{\left( 1-\cos ^{2}\theta _{%
\mathrm{\max }}\right) }=\sqrt{r_{1}^{2}-4\delta ^{2}}  \label{b5.22b}
\end{equation}

One can see, that the value $x_{\left( 2\right) }^{5}=\beta $ of the charge
coordinate $x^{5}$ of vector $\mathbf{P}_{1}\mathbf{P}_{2}$ is always less,
than $\frac{L}{2}+\frac{r_{1}}{2}$. Besides, the value $x_{\left( 2\right)
}^{5}=\beta $ of the charge coordinate $x_{\left( 2\right) }^{5}$ of vector $%
\mathbf{P}_{1}\mathbf{P}_{2}$ is less, than $L/2$, if the charge coordinate $%
x_{\left( 1\right) }^{5}=l$ of vector $\mathbf{P}_{0}\mathbf{P}_{1}$ is
larger, than $L/2$. In other words, the charge coordinate $x^{5}$ "reflects
itself" from the boundary $x^{5}=L/2$ in the following sense. If the
coordinate $x_{\left( 1\right) }^{5}$ of the vector $\mathbf{P}_{0}\mathbf{P}%
_{1}$ appears near the boundary $x^{5}=L/2$, $x_{\left( 1\right) }^{5}\in
\left( L/2-r_{1}/2,L/2\right) $, then coordinate $x_{\left( 2\right) }^{5}$
of the vector $\mathbf{P}_{1}\mathbf{P}_{2}$ may be larger than $L/2$. For
instance, it is possible, that $x_{\left( 2\right) }^{5}\in \left(
L/2,L/2+r_{1}/2\right) $. However, coordinate $x_{\left( 3\right) }^{5}$ of
the next link $\mathbf{P}_{2}\mathbf{P}_{3}$ will be less than $L/2$.
According to (\ref{b5.21}) $x_{\left( 3\right) }^{5}\in \left( \frac{L}{2}-%
\frac{3r_{1}}{2},\frac{L}{2}-\frac{r_{1}}{2}\right) $. Thus, the world chain
cannot go through the boundary $x^{5}=L/2$, although single points of the
world chain may have coordinate $x^{5}\in \left( L/2+r_{1}/2\right) $.
Behaviour of the world chain near the boundary $x^{5}=-L/2$ is the same, as
near the boundary $x^{5}=L/2$. The world chain reflects itself from the
boundary $x^{5}=-L/2$. Thus, the world chain will be placed in the region $%
-L/2<x^{5}<L/2$.

\label{2b}In the case of continuous space-time, when $\lambda _{0}=0$, the
world chain does not penetrate through the boundary $x^{5}=L/2$. In the case
of the discrete space-time, when $\lambda _{0}>0$, one point of the world
chain may penetrate through the boundary $x^{5}=L/2$. However, the next
point of the world chain returns to the region $-L/2<x^{5}<L/2$. We shall
refer to solutions, satisfying the condition (\ref{a5.10}), as basic
solutions, whereas the solutions, satisfying conditions (\ref{b2.47a}), as
additional solutions. \label{2e} Behaviour of the world chain near the
boundary $x^{5}=L/2$ reminds behavior of a quantum particle near the wall of
the potential hole.

Let us consider the case (additional solutions), when%
\begin{equation}
\gamma =-2L,\qquad L<l+\beta \leq 2L  \label{a5.21}
\end{equation}%
In the nonrelativistic case the equation (\ref{a5.9}) takes the form%
\begin{equation}
\mathbf{\alpha }^{2}+\left( \beta -l+2L\right) ^{2}=r_{2}^{2},\qquad r_{2}=%
\sqrt{8L^{2}-8lL+3\lambda _{0}^{2}}  \label{a5.22}
\end{equation}%
This equation describes the sphere of radius $r_{2}$ with center at the
point $\left\{ \mathbf{\alpha }_{\mathrm{c}}\beta _{\mathrm{c}}\right\}
=\left\{ \mathbf{0},l-2L\right\} $. It follows from (\ref{a5.22}) that%
\begin{eqnarray}
\alpha _{1} &=&r_{2}\sin \theta _{\mathrm{ad}}\sin \phi _{2}\sin \phi
_{3},\qquad \alpha _{2}=r_{2}\sin \theta _{\mathrm{ad}}\sin \phi _{2}\cos
\phi _{3},\qquad \alpha _{3}=r_{2}\sin \theta _{\mathrm{ad}}\cos \phi _{2},
\notag \\
\beta &=&l-2L+r_{2}\cos \theta _{\mathrm{ad}},\qquad \alpha _{0}=0,\qquad
\mathrm{for\ \ \ \ }r_{2}\cos \theta _{\mathrm{ad}}\geq 3L-2l  \label{a5.22b}
\end{eqnarray}%
where $\theta _{\mathrm{ad}},\phi _{2},\phi _{3}$ are arbitrary values. In
the case (\ref{a5.21}) and $\lambda _{0}=0$ we obtain, that $r_{2}=2L$ $%
\beta =l=L/2$.

Let us set
\begin{equation}
l=L/2+\delta ,\qquad \left\vert \delta \right\vert \leq \frac{r_{1}}{2}
\label{b5.23}
\end{equation}%
One obtains from the first inequality (\ref{a5.21}) and (\ref{a5.22b})%
\begin{equation}
0<2\delta -2L+r_{2}\cos \theta _{\mathrm{ad}}  \label{b5.25}
\end{equation}%
where $\theta _{\mathrm{ad}}$ is the angle between the axis and generatrix
of the cone with vertex at the point $\left( \mathbf{\alpha }_{\mathrm{c}%
}\beta _{\mathrm{c}}\right) =\left( \mathbf{0},l-2L\right) $. The cone is
based on the set of solutions on the sphere (\ref{a5.22}). One obtains from (%
\ref{b5.22b}) and (\ref{a5.21}) for the minimal value $\cos \theta _{\mathrm{%
\min }}$ of the quantity $\cos \theta _{\mathrm{ad}}$%
\begin{equation}
\cos ^{2}\theta _{\mathrm{\min }}=\frac{\left( 2L-2\delta \right) ^{2}}{%
4L^{2}-8L\delta +r_{1}^{2}}  \label{b5.25a}
\end{equation}%
\begin{equation}
\sin ^{2}\theta _{0}=1-\cos ^{2}\theta _{\mathrm{\min }}=\frac{\frac{1}{4}%
r_{1}^{2}-\delta ^{2}}{r_{2}^{2}},\qquad \delta ^{2}\leq \frac{1}{4}r_{1}^{2}
\label{b5.26}
\end{equation}

It follows from (\ref{b5.26}) that in the case (\ref{a5.21}) there are
additional solutions, if%
\begin{equation}
l=L/2+\delta ,\qquad \delta ^{2}<\frac{r_{1}^{2}}{4}\leq \frac{L^{2}}{4}
\label{b5.27}
\end{equation}%
The minimal value $\beta _{\min }$ is determined by the condition%
\begin{equation}
\beta _{\min }+l=L  \label{b5.27a}
\end{equation}%
which coincides with the maximal value (\ref{b5.22d}) for basic solutions.
Radius $R_{\mathrm{ad}}$ of the corresponding section of the sphere (\ref%
{a5.22}) has the form
\begin{equation}
R_{\mathrm{ad}}=r_{2}\sin \theta _{0}=\sqrt{r_{1}^{2}-4\delta ^{2}}
\label{b5.27b}
\end{equation}%
which coincides with the radius (\ref{b5.22b}). The solutions are shown in
figure 1

Action of the boundary $x^{5}=L/2$ changes the surface of the sphere of
basic solutions. The set of all solutions has a shape of two connected
spherical segments of different radius. Restriction (\ref{a2.44}) on the
electric charge of a particle is connected directly with a finite length of
the world chain links. In the discrete space-time geometry (\ref{a5.1}) the
length of the world chain link is finite by necessity (but not
infinitesimal), because the infinitesimal length does not exist. In the
continuous space-time geometry, the link of the world chain may be
infinitesimal, in principle. In this case the space-time compactification
does not restrict the maximal value of the electric charge.

Experimental data show, that the electric charge of a stable elementary
particle is equal to $0,\pm e_{0}$, where $e_{0}$ is the elementary charge
of an elementary particle.

\begin{figure}[ptb]
\begin{center}
\includegraphics [height=6cm,keepaspectratio]{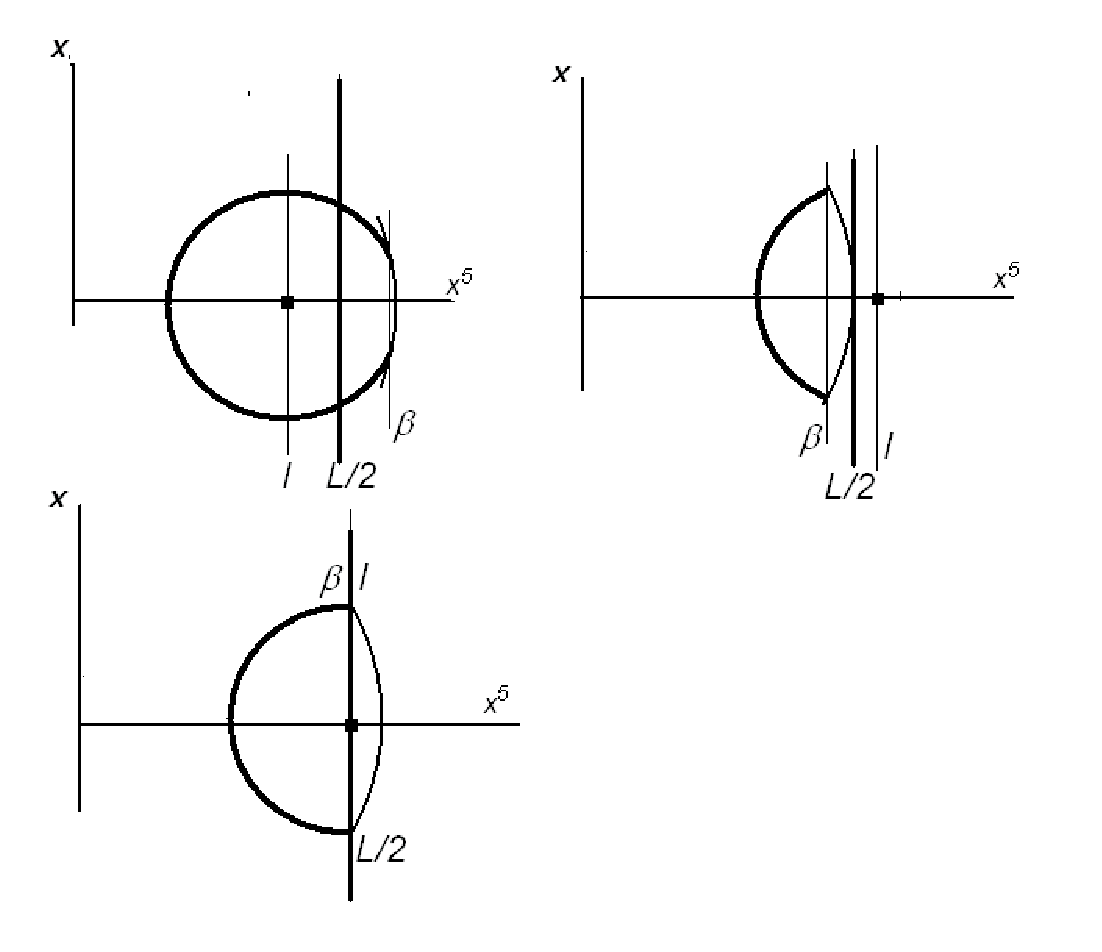}
\end{center}
\caption{ Set of solutions, depending on the value of $l$, shown by small
dark square. Usual solutions are shown by thick line. Additional solutions
are shown by thin line.}
\label{f1}
\end{figure}

\section{Restriction on maximal charge of elementary particle}

Thus, a compactification of the Kaluza-Klein space-time leads to a
discrimination of some values $p_{5}$ of the charge momentum component. The
discrimination is a corollary of the fact, that not all links of the world
chain of a particle are possible in the compactified geometry. In the
conventional approach to the Kaluza-Klein geometry, when the space-time
geometry is constructed as a Riemannian geometry (but not as a physical
geometry), there is no discrimination of the maximal value of the momentum
component $p_{5}$.

In all cases the electric charge has the form $e=ne_{0}$, where $e_{0}$ is
the elementary charge and $n$ is an integer number. This fact is connected
with the periodicity of the wave function with respect to fifth coordinate $%
x^{5}$. Averaging random world chains, one obtains a dynamic equation of the
type of the Schr\"{o}dinger equation. In the region, where the point $P_{2}$
of the vector $\mathbf{P}_{1}\mathbf{P}_{2}$ is placed on the sphere (\ref%
{a5.11}) this equation has the form%
\begin{equation}
i\hbar \frac{\partial \psi }{\partial t}=-\frac{\hbar ^{2}}{2m_{5}}\frac{%
\partial ^{2}\psi }{\partial \left( x^{5}\right) ^{2}}-\frac{\hbar ^{2}}{%
2m_{5}}\mathbf{\nabla }^{2}\psi  \label{a6.0}
\end{equation}%
In other regions, where the set of solutions is not a sphere, this equation
is modified. It takes a form of the Schr\"{o}dinger equation in a potential
hole with walls at $x^{5}=\pm L/2$. We shall not to take into account this
modification and use the equation (\ref{a6.0}) for approximate estimation of
connection between the period $2L$ and possible values of the of elementary
length $\lambda _{0}$.

Solution of equation (\ref{a6.0}) has the form
\begin{equation}
\psi \left( t,\mathbf{x},x^{5}\right) =\dsum\limits_{n}a_{n}\exp \left( -%
\frac{i}{\hbar }E_{n}t+i\frac{\mathbf{p}_{n}\mathbf{x}}{\hbar }+i\frac{%
\left( p_{5}\right) _{n}}{\hbar }x^{5}\right)  \label{a6.1}
\end{equation}%
where%
\begin{equation}
\left( p_{5}\right) _{n}=\frac{e}{\varkappa c}n,\qquad E_{n}=c\sqrt{%
m_{5}^{2}c^{2}+p_{n}^{2}+\left( p_{5}\right) _{n}^{2}}  \label{a6.2}
\end{equation}%
$n$ is an integer number, $\varkappa $ is some universal constant (\ref{a2.2}%
). $a_{n}$ are arbitrary complex numbers. The quantity $m_{5}=\mathrm{const}$
is a 5-mass of the particle, whereas the usual mass $m=\sqrt{%
m_{5}^{2}+p_{n}^{2}c^{-2}}$ depends on the fifth component $p_{5}$ of
momentum . In order that $\psi $ be a single-valued function of $x^{5}$, the
momentum $p_{5}$ is to have the form%
\begin{equation}
\left( p_{5}\right) _{n}=\frac{\pi \hbar }{L}k_{n},\qquad k_{n}\mathrm{\ \
is\ \ \func{integer}}  \label{a6.3}
\end{equation}

The wave function (\ref{a6.1}) must describe a stationary state of the
particle, because in a nonstationary state the charged particle, placed in a
potential hole, radiates electromagnetic waves. As a result the particle
appears very rapidly at a stationary state, where the charge density and
charge current are constant, and the particle ceases to radiate. The wave
function (\ref{a6.1}) is single-valued, if all $E_{n}$ in the sum (\ref{a6.1}%
) are equal, and any $E_{n}$ is not changed at a variation of $k_{n}$. These
conditions are fulfilled, if the sum (\ref{a6.1}) contains only one term,
and the momentum has the form%
\begin{equation}
p_{5}=\frac{\pi \hbar }{L}s\ \qquad \left( p_{5}\right) _{n}=\frac{e}{%
\varkappa c}n  \label{a6.4}
\end{equation}%
where $s$ is some definite integer number.

It follows from comparison of relations (\ref{a6.2}) and (\ref{a6.4}), that
\begin{equation}
\varkappa =\frac{e_{0}L}{\pi \hbar c}  \label{a6.5}
\end{equation}%
where $e_{0}$ is the elementary electrical charge, and $2L$ is the period of
the fifth coordinate $x^{5}$.

On the other hand, the momentum $p_{5}$ along the fifth direction is
connected with the geometrical momentum $\pi _{5}$ by means of the relation%
\begin{equation}
p_{5}=bc\pi _{5}  \label{a6.6}
\end{equation}%
According to the relation (\ref{a2.44}), where the universal constant $b$ is
defined by the relation (\ref{a3.4}),
\begin{equation}
\left\vert \pi _{5}\right\vert =\left\vert l\right\vert <\frac{L}{2},
\label{a6.7}
\end{equation}%
Using relations (\ref{a6.4}), (\ref{a6.6}), (\ref{a6.7}), one obtains
\begin{equation}
\qquad \pi _{5}=\frac{p_{5}}{bc}=\frac{1}{bc}\frac{\pi \hbar }{L}s=\pi \frac{%
\lambda _{0}^{2}}{L}s  \label{a6.8}
\end{equation}%
and%
\begin{equation}
\left\vert s\right\vert <\frac{L^{2}}{2\pi \lambda _{0}^{2}}  \label{a6.9}
\end{equation}%
Taking maximal value $r_{1}=\sqrt{3}\lambda _{0}=L/2$, one obtains%
\begin{equation}
\left\vert s\right\vert <\frac{3L^{2}4}{2\pi L^{2}}=1.\,\allowbreak 909\,9
\label{a6.10}
\end{equation}
$\left\vert s\right\vert =1$, and the module $\left\vert e\right\vert $ of
the charge $e$ of a stable elementary particle is not more than the
elementary charge $e_{0}$. In general, the approximation (\ref{a6.0}) is too
rough, and the relation (\ref{a6.10}) may not be considered as a true
relation. Firstly we have used the approximate equation (\ref{a6.1}).
Secondly, the choice $r_{1}=L/2$ is not founded exactly. However, it is
important, that in any case the maximal electric charge of a stable
elementary particle is restricted. The exact value of $s$ may be obtained at
at proper choice of $\lambda _{0}$. The relation (\ref{a6.10}) shows, that
if the period $2L$ of the fifth coordinate $x^{5}$ is of the order of $%
\lambda _{0}$, it is possible such an interrelation between $L$ and $\lambda
_{0}$, that the module $\left\vert e\right\vert $ of the charge $e$ of a
stable elementary particle is not more, than the elementary charge $e_{0}$.
It is essential, that the restriction has a geometrical form. It connects
the elementary length $\lambda _{0}$ with the length $2L$ of
compactification.

Thus, compactification of the Kaluza-Klein space-time geometry imposes
restrictions on possible values of the electric charge of an elementary
particle. One needs only to use the physical geometry, which uses uniform
formalism for description of continuous and discrete geometries.

\section{Concluding remarks}

Discreteness of the space-time in microcosm seems to be a more simple and
reasonable supposition, than the opposite supposition on continuous
space-time equipped by quantum principles. Discreteness of the space-time
admits one to describe quantum effects without referring to quantum
principles. Describing discreteness of the space-time, the elementary length
$\lambda _{0}$ determines the quantum constant $\hbar $. The space-time
discreteness appears to be compatible with its isotropy and its uniformity.
However, this compatibility can be understood only in framework of physical
geometry, which uses the same formalism for description of discrete and
continuous geometries. Combination of the discrete space-time with its
compactification admits one to obtain restrictions on the electric charge of
stable elementary particles. These restrictions are known from experiments,
but they have no explanation in the framework of the conventional quantum
theory.

\end{document}